\documentclass[reprint,superscriptaddress,amsmath,amssymb,aps,pra]{revtex4-1}

\usepackage{graphicx}
\usepackage{bm}
\usepackage{color}
\usepackage[normalem]{ulem} 

\begin{document}


\title{Intensity modulation as a preemptive measure against blinding\\of single-photon detectors based on self-differencing cancellation}

\author{Alexander Koehler-Sidki}
\affiliation{Toshiba Research Europe Ltd, Cambridge Research Laboratory, 208 Cambridge Science Park, Milton Road, Cambridge, CB4 0GZ, United Kingdom}
\affiliation{Engineering Department, University of Cambridge, 9 J. J. Thomson Avenue, Cambridge, CB3 0FA, United Kingdom}
\author{Marco Lucamarini}
\affiliation{Toshiba Research Europe Ltd, Cambridge Research Laboratory, 208 Cambridge Science Park, Milton Road, Cambridge, CB4 0GZ, United Kingdom}
\author{James F. Dynes}
\affiliation{Toshiba Research Europe Ltd, Cambridge Research Laboratory, 208 Cambridge Science Park, Milton Road, Cambridge, CB4 0GZ, United Kingdom}

\author{George L. Roberts}
\affiliation{Toshiba Research Europe Ltd, Cambridge Research Laboratory, 208 Cambridge Science Park, Milton Road, Cambridge, CB4 0GZ, United Kingdom}
\affiliation{Engineering Department, University of Cambridge, 9 J. J. Thomson Avenue, Cambridge, CB3 0FA, United Kingdom}%
\author{Andrew W. Sharpe}
%
%
%
\author{Zhiliang Yuan}
\email{zhiliang.yuan@crl.toshiba.co.uk}
\author{Andrew J. Shields}
\affiliation{Toshiba Research Europe Ltd, Cambridge Research Laboratory, 208 Cambridge Science Park, Milton Road, Cambridge, CB4 0GZ, United Kingdom}%

\date{\today}

\begin{abstract}
\noindent Quantum key distribution is rising as an important cryptographic primitive for protecting the communication infrastructure in the digital era.
However, its implementation security is often weakened by components whose behavior deviates from what is expected. Here, we analyse the response of a self-differencing avalanche photodiode, a key enabler for high speed quantum key distribution, to intense light shone from a continuous-wave laser. Under incorrect settings, the cancellation entailed by the self-differencing circuitry can make the detector insensitive to single photons. However, we experimentally demonstrate that even in such cases intensity modulation can be used as an effective measure to restore the detector's expected response to the input light.
%
\end{abstract}


\maketitle


\section{Introduction}
\label{sec:level1}

\noindent Quantum key distribution (QKD) promises information theoretic security that is guaranteed by the laws of quantum mechanics \cite{BB14}.
Its potential as a cryptographic primitive has stimulated significant developments in recent years and has resulted in pilot network field trials in several continents \cite{PPA+09,sasaki11,yang_china_2017,Wang:14,Mirza:10,Batt14,QKD17}. However, actual components in QKD implementations can deviate from their ideal behaviour, 
creating side-channels that might be exploited by an eavesdropper (Eve), thus threatening the theoretical security promised by QKD \cite{lydersen_hacking_2010_pra,gerhardt_full-field_2011,Li11,Lyder11,Wiech11,Meda17}. 

To mitigate this security risk, active or passive countermeasures can be considered. 
In an active approach, the legitimate QKD users (Alice and Bob) monitor in real time the device parameters that change under Eve's attack \cite{Silva12, maroy_2017}. In a passive approach, they add extra guarding components to thaw Eve's attempt \cite{lucamarini_tha_2015, lee_countermeasure_2016}. Using the Trojan-horse attack \cite{vakhitov_tha_2011,gisin_tha_2006} as an example,  Alice and Bob can employ either a watchdog detector to actively detect Eve's presence \cite{stucki_pandp_2002} or a combination of an optical filter, attenuator and isolator to passively prevent Eve's light from reaching the encoding devices \cite{lucamarini_tha_2015}. 

Single photon detectors are the most vulnerable component in a QKD system, especially when inappropriately operated \cite{yuan_resilience_2011_pra,AKS2017SPIE}, due to their optical exposure to Eve through the quantum channel. Those based on semiconductor avalanche photodiodes (APDs) can be attacked using a strong optical signal in the so-called ``blinding attack'' \cite{lydersen_hacking_2010_pra,gerhardt_full-field_2011}. The severity of this attack has motivated a number of active countermeasures, including monitoring the APD current \cite{yuan_resilience_2011_pra}, varying \cite{lim_random_2015, huang_testing_2016} or calibrating \cite{maroy_2017} in real time its detection efficiency, or using a watchdog detector \cite{sajeed_watchdogdet_2015}. Extraordinarily,  detector side-channel free QKD \cite{lo12,braunstein12} closes all side channels in the measurement devices, but requires significant changes in users' apparatus and does not offer an easy upgrade to existing systems.

Self-differencing (SD) InGaAs APDs are an important class of detectors for high bit rate QKD systems \cite{dynes16,Yuan18}. They enable detection of extremely weak signals using a passive circuit for cancellation of the intense background capacitive response \cite{yuan_highspeedir_2007_pra}, thus supporting count rates in the GHz range \cite{patel12a}, detection efficiency up to 55\% \cite{comandar_gigahertz-gated_2015_pra}, room-temperature operation \cite{Comandar:rm_temp_det_pra} and superior resilience to background noise photons \cite{patel12prx}. On the other hand,
the self-cancellation nature of the SD circuit prevents the detection of consecutive identical signals \cite{dynes08}. This might be exploited by Eve to perform the blinding attack with a lower optical power \cite{jiang_intrinsic_2013_pra}, especially for certain settings of the detector \cite{AKS2017SPIE}.

Here, we propose 
taking preemptive action against blinding by placing
an intensity modulator (IM) in front of the receiver's measurement apparatus. The use of low extinction ratio modulation will not severely attenuate the incoming quantum signal, but the IM will create sufficient modulation in the detector's photocurrent, which is detectable by SD circuitry and discrimination electronics. We experimentally demonstrate this method on an SD InGaAs APD.

This idea is similar in spirit to the random variation of an APD's detection efficiency, suggested and partially implemented in \cite{lim_random_2015}, which was shown to be ineffective against a refined Eve's attack \cite{huang_testing_2016}. However, our proposal contains some notable differences. An APD endowed with an SD circuit sets a more challenging target to the eavesdropper. In fact, Eve has to send sequential light pulses with identical intensity to cause blinding. Any small deviation from this condition is likely to cause a detection event with an associated 50\% quantum bit error rate (QBER). At the same time, a long sequence of bright optical pulses generates a high photocurrent which is easily detectable \cite{yuan_avoiding_2010_pra}. When an IM with random modulation is added on top of this already compelling situation, the constraints for Eve become extremely stringent. In particular, we found no room for blinding in our experiment.



\section{Blinding attacks and countermeasures}

Single photon sensitivity of an APD relies on having an electrical excess bias above its breakdown voltage to enable avalanche multiplication of a single photo-carrier. In the blinding attacks, Eve erodes this excess by inducing an electrical current flowing through the biasing circuit \cite{yuan_resilience_2011_pra} or heating up the device to raise its breakdown voltage \cite{thermal_Lydersen:10}. These attacks are realised through injecting a strong laser signal into Bob's module through the quantum channel and making both of his detectors blind, i.e., insensitive to single photons.
At this point, Eve performs a modified intercept-resend attack to take control of Bob's detectors. She measures the state prepared by Alice and re-sends a suitably prepared faked state encoded in a strong optical pulse. This will then trigger a detector count only when Bob chooses a measurement basis that is identical to Eve's. In this attack, Eve can gain full information about the final key \cite{fakedstate_2005}.

\begin{figure}
  \centering
  \includegraphics[width=0.46\textwidth]{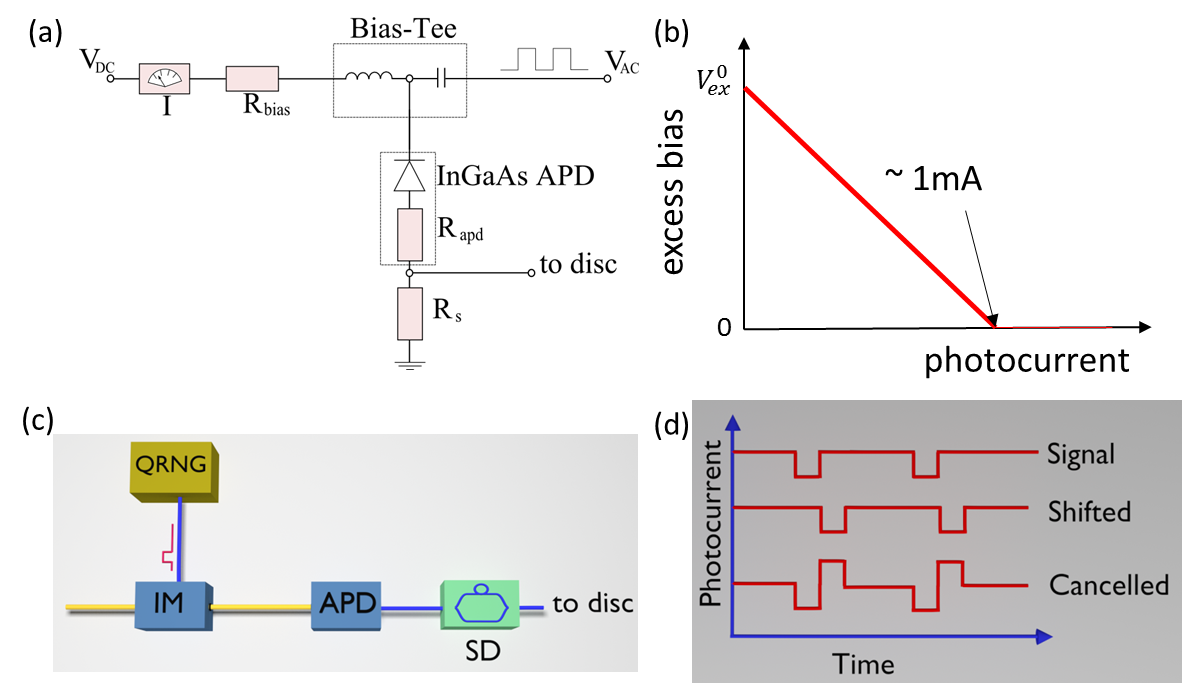}
  \caption{ (a) Schematic of a biasing scheme for a gated APD. $V_{DC}$ : DC bias component; $V_{AC}$: AC bias component; $R_{\textrm{bias}}$: biasing resistor; $R_{\textrm{apd}}$: APD  internal resistance; $R_{\textrm{s}}$: sensing resistor. (b) Reduction in the excess bias due to photocurrent. (c) Schematic of the measure against blinding. IM: intensity modulator; QRNG: quantum random number generator; SD: Self-Differencer. (d) Effect of the intensity modulation on the SD photocurrent in presence of bright light inputted by Eve.}
  \label{fig:countermeasures}
\end{figure}

Consider an APD detector operated in Geiger mode with an excess voltage of $V_{ex}^0$, as shown in Fig. \ref{fig:countermeasures}(a).
To completely erode this excess and blind the detector, Eve needs to create a photocurrent $I$ that can be approximated as
\begin{align}
I = \frac{V_{\textrm{ex}}^0}{R_{\textrm{bias}} + R_{\textrm{apd}} + R_{\textrm{s}}/2},
\label{eq:circuit}
\end{align}
where $R_{\textrm{bias}}$, $R_{\textrm{apd}}$ and $R_{\textrm{s}}$ are the resistance values for the biasing resistor, the APD itself and the sensing resistor, respectively. In a usual setup, $R_{\textrm{bias}} = 0$ and the current is determined mainly by the value of $R_{\textrm{apd}}$. Its typical value is on the order of 1~mA, see Fig.~\ref{fig:countermeasures}(b).  This large current, together with the gain modulation effect by the detector gating, has previously enabled gated-APDs to avoid the blinding attack when their discrimination levels are appropriately set \cite{yuan_resilience_2011_pra}.



Here we propose a different measure, schematically shown in Fig.~\ref{fig:countermeasures}(c). We insert an IM, driven by a quantum random number generator (QRNG), in front of the optical fibre input of the APD detector and an SD circuit after its electrical signal output. The SD circuit splits the APD output into two equal components, delays one of them, and then combines the two signals differentially, see Fig.~\ref{fig:countermeasures}(d). The positive peaks of the resulting photocurrent can then be detected by the discrimination electronics.

The IM acts as an optical shutter and stops any incoming light for a short duration at random times. Under normal conditions, \textit{i.e.}, in the absence of Eve, this would cause a decrease in the counts seen by Bob every time the IM is activated.
Correspondingly, the resulting avalanche current would exhibit a waveform containing a positive current peak followed by a negative dip. On the contrary, if Eve sends her blinding pulses into Bob's module, the IM's activation would increase the counts seen by Bob, due to the SD effect, and the resulting photocurrent would exhibit a negative current dip followed by a positive peak. Because this outcome is distinctively different from that under normal conditions, it represents a clear signature of Eve's presence.

Even without correlating Bob's counts to the IM's activation times, the presence of the IM and SD circuitry make it possible to restore the APD's count rate and prevent its blinding. This is a simple consequence of the fact that, irrespective of the polarity of the photocurrent, the positive peak is always well above the detector's discrimination level, for a detector that has been correctly set up. So, for simplicity, we decided to not take advantage in this work of the ``fine-grained'' signatures based on correlating the counts with the IM or based on the avalanche's polarity and focus rather on the ``coarse-grained'' signature represented by the APD's counts. The analysis of the whole statistics available to Bob is left for future studies and can only reinforce the results presented here.



\section{Experimental setup}

\begin{figure}
  \centering
  \includegraphics[width=0.47\textwidth]{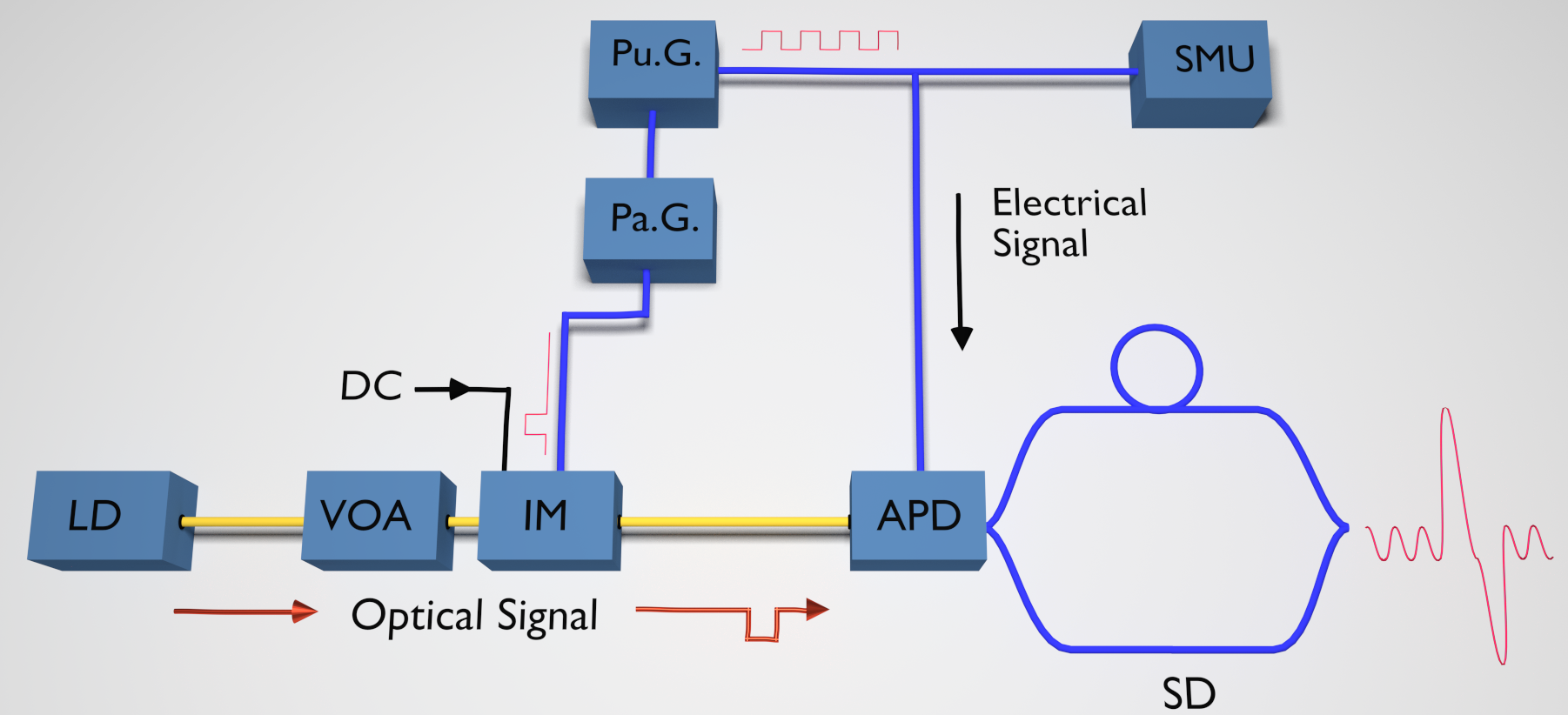}
  \caption{Experimental setup. Pu.G.: Pulse Generator; Pa.G.: Pattern Generator; SMU: Source Measure Unit; LD: Laser Diode; VOA: Variable Optical Attenuator; IM: Intensity Modulator; APD: Avalanche Photodiode; SD: Self-Differencer.}
  \label{fig:exp_setup}
\end{figure}

To investigate the efficacy of the proposed measure, we adopt the experimental setup shown in Fig.~\ref{fig:exp_setup}, which includes both an IM and an SD circuit. An InGaAs/InP APD is thermo-electrically cooled to  --30 $^{\circ}$C and operated with a constant DC bias of 51.6~V and a 1~GHz square-wave signal with a peak-to-peak amplitude of 4.6 V. A telecom C-band continuous wave laser diode emitting polarised light is used to illuminate the APD.
The APD resistance is measured to be $R_{\textrm{apd}}=1$~k$\Omega$, and no biasing resistor is used, $R_{\textrm{bias}}=0$.  We use a variable optical attenuator to provide a 120~dB intensity variation range and a LiNbO$_{3}$ intensity modulator driven by a pattern generator to modulate the optical power. A high-bandwidth oscilloscope (16~GHz) is used to analyze the SD output. We determine the appropriate discrimination level for the SD-APD to be 18~mV and measure a detection efficiency of 26\% for pulsed light and a dark count rate of 23~kHz at this level. We note that under this operation condition it is not possible for Eve to blind the detector using continuous-wave illumination because the increasing APD capacitive response is sufficient to counter Eve's blinding effort \cite{KoehlerSidki18}.

\section{Effect of the intensity modulation on the count rate}

\begin{figure}
  \centering
  \includegraphics[width=0.46\textwidth]{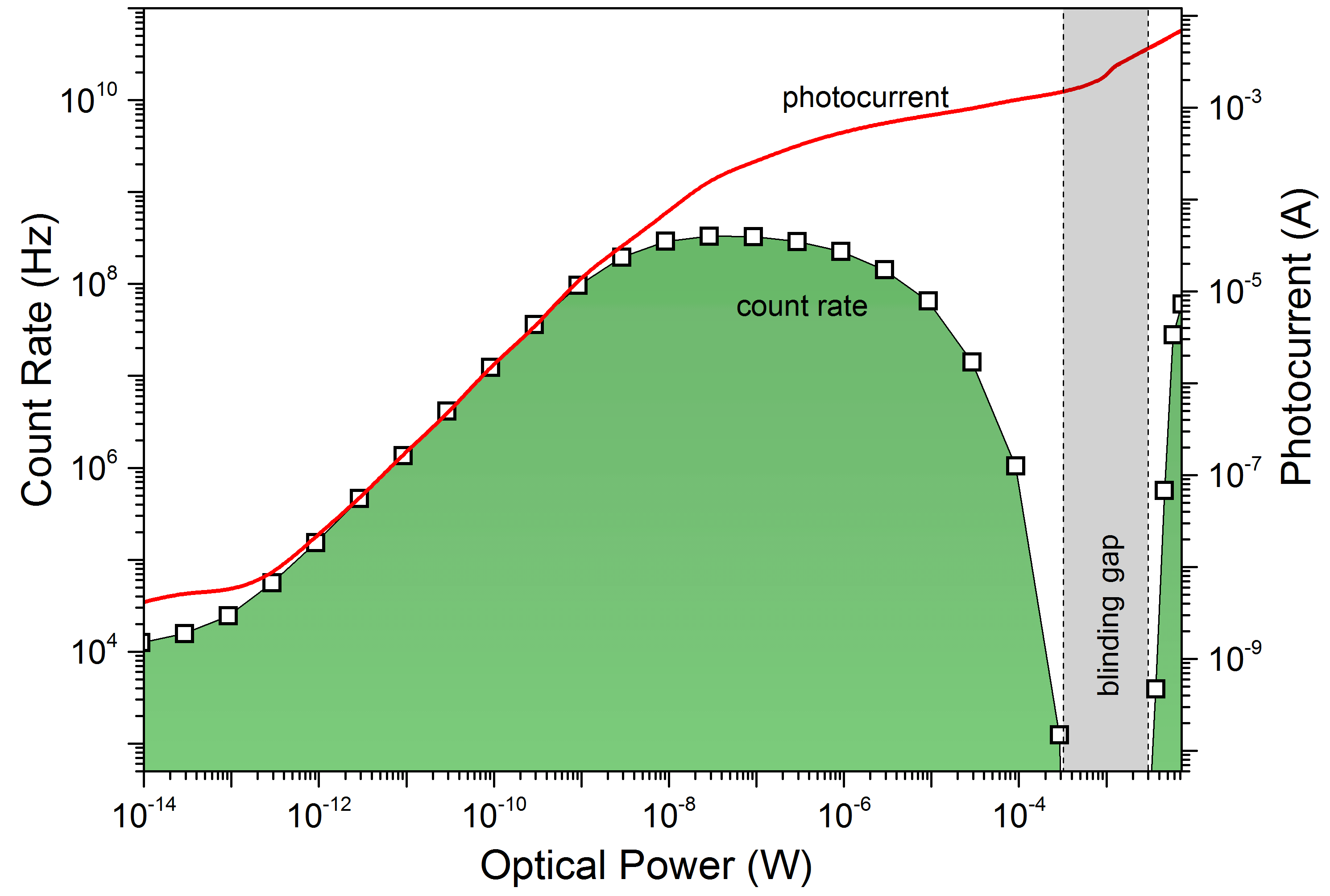}
  \caption{The detector count rate in the case of an inappropriately high discrimination level and associated photocurrent as a function of the incident optical power. Count rate and photocurrent can be simultaneously measured and pose stringent constraints on Eve's actions.}
  \label{fig:CR}
\end{figure}

We first demonstrate an experimental condition under which the SD-APD can be blinded.  By deliberately setting the discrimination level inappropriately high, at 26~mV, we measure the count rate of the SD-APD as a function of incident optical power measured directly after the IM. Here, the RF input to the IM is disabled and its DC bias is adjusted to have a maximum transmission. Figure~\ref{fig:CR} shows the count rate and photocurrent as a function of the incident optical power.  The detector exhibits a blinding gap between 300~$\mu$W and 3~mW, within which the detector count rate falls to zero. Such a blinding gap enables Eve to gain complete control of the detector. The photocurrent follows the count rate and grows linearly before the count rate saturation, and then grows sub-linearly (100~nW -- 1$\mu$W) before becoming quasi-linear with the incident optical power ($>1\mu$W). In the blinding gap, the photocurrent is measured to exceed 1~mA.

Such a large photocurrent offers an opportunity to close the blinding gap by modulating the intensity of the attacking signal.
 As illustrated earlier with Fig.~\ref{fig:countermeasures}(d), an intensity modulation creates a dip in the photocurrent. The SD circuit converts each dip to a positive peak which can trigger the detector discrimination circuit when there is sufficient modulation depth.
A random pattern produces photocurrent dips at a rate that is 1/4 of the signal clock rate. For simplicity we simulate this rate in our experiment by applying an RF signal to the IM using a repetitive modulation pattern ``0001", which we label as ``$1/4$" modulation.  We set the RF amplitude to 4~V, achieving half-wave modulation and an intensity extinction ratio of 23~dB.
This pattern carves a 1~ns hole for every 4 ns duration in the attacking light intensity.

\begin{figure}
  \centering
  \includegraphics[width=0.46\textwidth]{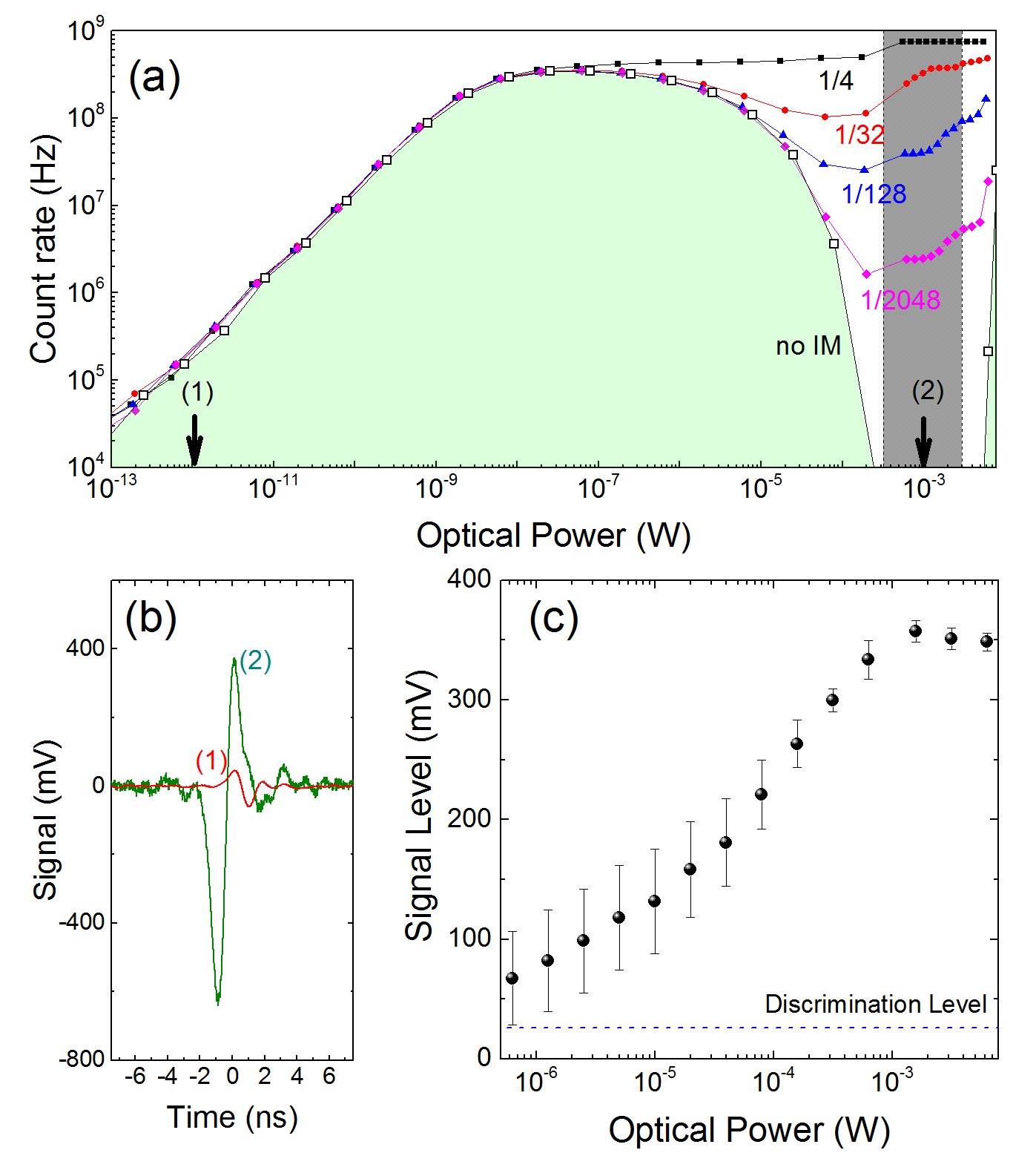}
  \caption{(a) APD count rates as a function of incident optical power with different modulations applied to the intensity modulator. An RF amplitude of 4~V is used to produce half-wave modulation and a modulation extinction ratio of 23~dB. (b) The SD output recorded by the oscilloscope at points (1) and (2) in (a) with the attacking laser being modulated by a ``1/32" pattern. (c) Signal level of the main positive peak as a function of optical power. }
  \label{fig:im}
\end{figure}

Using the ill-set discrimination level of 26~mV, we measure the count rate versus the incident optical power with the result shown as black solid circles in Fig.~\ref{fig:im}(a). The intensity modulation causes distinctively different count rate behavior for higher optical power when compared with the case without intensity modulation (open squares). The count rate stays above 250~MHz from 100~nW to 7.5~mW, without any sign of it falling. Despite the high discrimination level, the IM successfully removes the former SD-APD's blinding gap.

We attribute the closure of the blinding gap to the applied intensity modulation. To illustrate this, we compare two SD-APD output waveforms recorded under vastly different optical powers.  In the single photon counting regime, the APD produces a positive, current spike and therefore its SD output becomes a positive spike followed by its negative copy 1~ns afterwards, see waveform 1 in Fig.~\ref{fig:im} (b).  With an optical power of 1~mW, the SD-APD output waveform reverses its polarity (waveform 2) because the intensity modulation carves a hole in the photocurrent, instead of a current spike for a single-photon induced avalanche. The signal level is about 9 times as strong as the single-photon induced avalanche, and can therefore overcome the detector discrimination level.  This confirms the intuition given at the beginning of the paper.
For the case of 1/4 modulation, the count rate saturates close to 750 MHz, which can be explained by having two ripples
after the main avalanche peak in Fig.~\ref{fig:im}(b) overcoming the discrimination threshold as well.

Figure~\ref{fig:im}(c) plots the signal level of the main positive peak induced by the IM as a function of incident optical power.  Over the incident power spanning over 4 orders of magnitude between 0.7~$\mu$W and 7~mW, the IM induced signal has a significant margin to overcome the discrimination level, even though it was inappropriately set. At an optical power of $1~\mu W$, where the count rate with no IM (open squares in Fig.~\ref{fig:im}(a)) begins to fall, the signal level in Fig.~\ref{fig:im}(c) is over 50~mV and continues to increase in amplitude to over 300~mV at an optical power of greater than 1~mW. Within the power range Eve needs for blinding, each intensity modulation is guaranteed to produce at least one detector count. This is in agreement with our experiment using sparser modulation patterns, as shown in Fig.~\ref{fig:im}(a). A sparser modulation results in a proportionally lower bottom-out count rate in the blinding gap.

\begin{figure}[htbp!]
  \centering
  \includegraphics[width=0.46\textwidth]{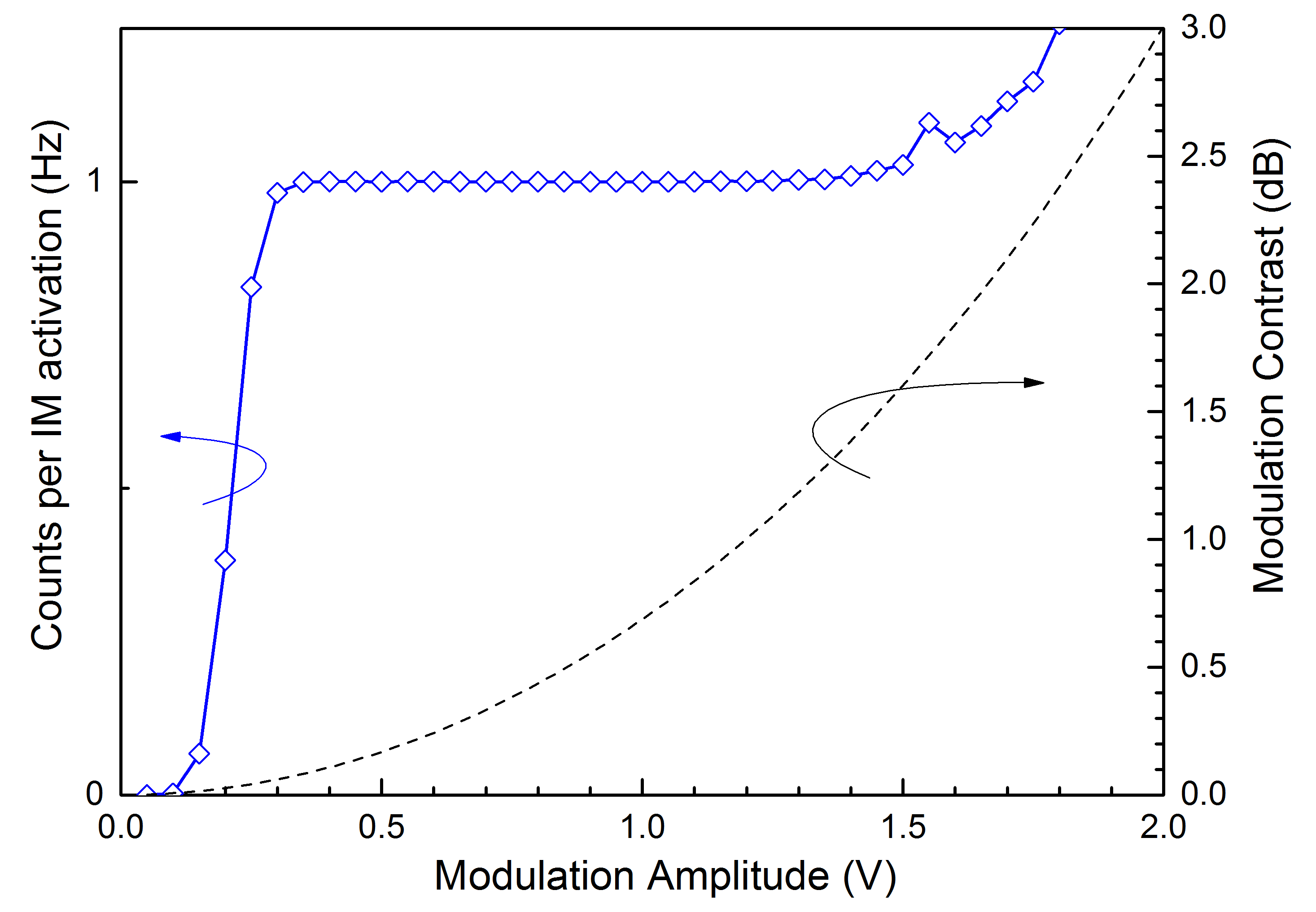}
  \caption{APD counts per IM activation and IM contrast as a function of the AC signal driving the IM, for a constant incident optical power of 1 mW and a modulation pattern 1/128.  We note that the APD discrimination level is deliberately set too high (26~mV) to enable blinding when the IM is switched off.}
  \label{fig:Opt_IM}
\end{figure}

The significant margin in the signal level shown in Fig.~\ref{fig:im}(c) offers room to relax the requirement on the IM's modulation contrast, thus minimising the loss that the IM would introduce.
We determine the lowest modulation contrast by measuring the count rate probability as a function of the RF signal amplitude applied to the IM with the DC set to maximum transmission. Here, the modulation frequency is 1/128 of 1~GHz and the incident optical power is 1~mW. As shown in Fig.~\ref{fig:Opt_IM}, a modulation signal with amplitude 0.3~V can always induce at least one detector count. More counts are possible due to the ripples in the output waveform also overcoming the discrimination threshold. This RF level corresponds to an intensity contrast of  0.06~dB. The number of counts increases above unity at a modulation amplitude higher than 1.5~V because the amplitude of the signal ripple (see Fig.~\ref{fig:im}(b)) rises above the discrimination level.

\section{Intensity modulation to prevent blinding}

%

In this section we discuss how the IM represents a potential countermeasure to the blinding attack. In doing so, we exclude side effects due to imperfect electronics or to a wrong setting of the parameters. We also neglect any artificial additions used to facilitate post-processing or software implemention, such as dead time, which may be exploited by Eve with a modified attack \cite{Wiech11}. We assume that the IM is driven by a true random generator, so Eve cannot deterministically predict the modulation effect and prepare her blinding pulses accordingly. We also consider a sufficient modulation depth to ensure a strong signal difference between modulated and unmodulated gates, which guarantees the occurrence of a detector count, as we have shown. The synchronization of the IM pulses, as well as the differential delay of the SD circuit, must be carefully chosen such that the counts by IM activation fall within the acceptance time window of the QKD system. Finally, a spectral filter should be applied in the QKD receiver to limit Eve's choice of wavelength and ensure that the modulation effect on Eve's attacking signal takes place.

The main security observation is that each IM-induced electrical signal overcomes the discrimination level and deterministically generates at least one detector count. These counts have equal probabilities of contributing a correct or incorrect bit in the sifted key, thus generating an overall 50\% QBER. Therefore we can choose the probability to activate the IM, $P_{IM}$, such that the resulting QBER in the presence of Eve exceeds the security tolerance of the protocol. If this happens, the protocol is aborted and no insecure key is distilled.

To decide the correct value for $P_{IM}$, let us consider the BB84 protocol~\cite{BB14}, which features a security tolerance of 11\%. Suppose that
$P_{IM}=25\%$. In this case, we have a guarantee that if Eve always launches her attack, Bob will see at least 1 click in 4 input pulses, which would cause a QBER, $Q_b$, equal to or larger than 12.5\%. In this case the key rate is zero, $R(Q_b)=0$. If Eve works in ``burst-mode'', she cannot do better than in the previous scenario. On the $N$ preparations effected by Alice, she could intercept-and-resend $N_b$ consecutive blinding signals in bursts and mount her attack only on these bursts, while blocking the remaining $N-N_b$ quantum signals. Even in this case, the IM would guarantee the final QBER to be above 12.5\%, causing zero key rate. Moreover, at the beginning of the train of blinding pulses prepared by Eve, the SD effect would cause one additional count in Bob's detectors, making this scenario more favourable to the legitimate users.

On the other hand, Eve could let a fraction of Alice's signals pass undisturbed. In this case, Eve gets no information on the undisturbed signals, but the resulting QBER would be smaller than the protocol's tolerance and the users would not abort the transmission. However this case is still secure due to the fact that the key rate is a convex function of the QBER (see e.g.~\cite{Koa06}).
Suppose that $C_b$ and $C_u$ ($Q_b$ and $Q_u$) are the count rates (QBERs) pertaining to blinding and undisturbed pulses, respectively. Then the average QBER seen by the users is $Q=C_b Q_b+C_u Q_u$. The convexity of the key rate and the fact that $R(Q_b)=0$ imply that
\begin{equation}\label{eq:conv1}
  R(Q)\leq R(Q_b)+R(Q_u)=R(Q_u),
\end{equation}
where $R(Q_b)$ and $R(Q_u)$ are the key rates from separate blinding and undisturbed pulses, respectively. Eq.~\eqref{eq:conv1} shows that there is at least a fraction $R(Q_u)$ of secure bits in the users' signals. This comes from the fact that $R(Q_u)$ is associated with the undisturbed pulses.  In the real case, the users measure $Q$ and distill a secure fraction $R(Q)$. This, by virtue of Eq.~\eqref{eq:conv1}, is a pessimistic estimate of the fraction of secure bits in the sample, hence the protocol is secure.

\section{Summary and Discussion}

In conclusion, we have devised and experimentally demonstrated a new technique to mitigate detector blinding.
By using an intensity modulator and an SD circuit, we modulated the incoming light to create uneven avalanches for the case of strong input light. Significantly, whilst this protects the detector from Eve it only introduces a small intrinsic attenuation of Alice's signal. We showed that a modulation depth of 0.06~dB is sufficient to prevent an SD detector from being blinded. In our experimental test, we adopted a continuous-wave laser. A pulsed laser would be no more effective at blinding an SD detector as it creates more intensity fluctuations, to which the detector is very sensitive.
Although intensity modulation to prevent Eve's faked-state attack has been previously addressed in the literature \cite{huang_testing_2016}, this was concerned with controlling an \textit{already blinded} detector. Our approach, on the contrary, includes a SD circuit to prevent blinding in the first place and thus eliminate the possibility of a faked-state attack at the root.

The proposed IM measure entails a considerable extrinsic loss penalty of around 2.5--5~dB, arising from imperfect intensity modulators based on LiNbO$_3$ and will therefore negatively impact the secure key rate. The loss associated with the product of the modulation rate of 1/4 and extinction ratio of 0.06 dB is comparatively negligible, hence the key rate in the presence of modulation would be $0.315$ times the unmodulated key rate and the distance would be shortened by 25 km, assuming a maximum insertion loss of 5 dB. Although QKD systems typically have two or more detectors, placing an intensity modulator in front of only one would be sufficient to demonstrate the presence of Eve. We note that existing modulators were designed to achieve high modulation depth which requires a lengthy crystal waveguide for electro-optical interaction. With a reduced modulation depth, the insertion loss can be made significantly smaller, thereby alleviating the loss penalty. Although intensity modulators are typically polarization sensitive, Eve cannot mount an attack such that she simply sends light of a polarization which does not experience modulation. This is because the detectors in a QKD system always see a fixed polarization, whether in phase-encoded schemes such as \cite{dixon_2017}, which contain an electronic polarization controller followed by a polarizing beamsplitter (PBS) or in polarization-encoded schemes, such as in \cite{gerhardt_full-field_2011}, which also have PBSs before the detectors. The use of an IM also requires a random number generator (RNG). Since Bob typically already has an RNG for the purpose of active basis selection, he can use the same RNG operated at twice the clock rate for the IM, which would not open additional side-channels.

\section*{Acknowledgements}
A. K.-S. gratefully acknowledges financial support from the EPSRC and Toshiba Research Europe Ltd.
\bibliography{apssamp}

\providecommand{\noopsort}[1]{}\providecommand{\singleletter}[1]{#1}%
\begin{thebibliography}{10}
\providecommand{\url}[1]{\texttt{#1}}
\providecommand{\urlprefix}{URL }
\providecommand{\eprint}[2][]{\url{#2}}

\bibitem{BB14}
C.~H. Bennett and G.~Brassard.
\newblock Theor. Comput. Sci. \textbf{560}, 7 (2014).

\bibitem{PPA+09}
M.~Peev, C.~Pacher, R.~All{\'{e}}aume, C.~Barreiro, J.~Bouda, W.~Boxleitner
  \emph{et~al.}
\newblock New J. Phys. \textbf{11}, 075001 (2009).

\bibitem{sasaki11}
M.~Sasaki, M.~Fujiwara, H.~Ishizuka, W.~Klaus, K.~Wakui, M.~Takeoka
  \emph{et~al.}
\newblock Opt. Express \textbf{19}, 10387 (2011).

\bibitem{yang_china_2017}
Y.~Yang.
\newblock China trial paves way for ‘unhackable’ communications network
  (2017).

\bibitem{Wang:14}
S.~Wang, W.~Chen, Z.-Q. Yin, H.-W. Li, D.-Y. He, Y.-H. Li \emph{et~al.}
\newblock Opt. Express \textbf{22}, 21739 (2014).

\bibitem{Mirza:10}
A.~Mirza and F.~Petruccione.
\newblock J. Opt. Soc. Am. B \textbf{27}, A185 (2010).

\bibitem{Batt14}
D.~Hayford.
\newblock \emph{Quantum-based Product Development at {Battelle}}.

\bibitem{QKD17}
D.~Bunandar, A.~Lentine, C.~Lee, H.~Cai, C.~M. Long, N.~Boynton \emph{et~al.}
\newblock Phys. Rev. X \textbf{8}, 021009 (2018).

\bibitem{lydersen_hacking_2010_pra}
L.~Lydersen, C.~Wiechers, C.~Wittmann, D.~Elser, J.~Skaar, and V.~Makarov.
\newblock Nat. Photonics \textbf{4}, 686 (2010).

\bibitem{gerhardt_full-field_2011}
I.~Gerhardt, Q.~Liu, A.~Lamas-Linares, J.~Skaar, C.~Kurtsiefer, and V.~Makarov.
\newblock Nat. Commun. \textbf{2}, 349 (2011).

\bibitem{Li11}
H.-W. Li, S.~Wang, J.-Z. Huang, W.~Chen, Z.-Q. Yin, F.-Y. Li \emph{et~al.}
\newblock Phys. Rev. A \textbf{84}, 062308 (2011).

\bibitem{Lyder11}
L.~Lydersen, N.~Jain, C.~Wittmann, O.~Mar\o{}y, J.~Skaar, C.~Marquardt
  \emph{et~al.}
\newblock Phys. Rev. A \textbf{84}, 032320 (2011).

\bibitem{Wiech11}
C.~Wiechers, L.~Lydersen, C.~Wittmann, D.~Elser, J.~Skaar, C.~Marquardt
  \emph{et~al.}
\newblock New Journal of Physics \textbf{13}, 013043 (2011).

\bibitem{Meda17}
A.~Meda, I.~P. Degiovanni, A.~Tosi, Z.~Yuan, G.~Brida, and M.~Genovese.
\newblock Light: Science and Applications \textbf{6}, e16261 (2017).

\bibitem{Silva12}
T.~F. da~Silva, G.~B. Xavier, G.~P. Temporao, and J.~P. von~der Weid.
\newblock Opt. Express \textbf{20}, 18911 (2012).

\bibitem{maroy_2017}
{{\O}}ystein Mar{{\o}}y, V.~Makarov, and J.~Skaar.
\newblock Quantum Science and Technology \textbf{2}, 044013 (2017).

\bibitem{lucamarini_tha_2015}
M.~Lucamarini, I.~Choi, M.~B. Ward, J.~F. Dynes, Z.~L. Yuan, and A.~J. Shields.
\newblock Phys. Rev. X \textbf{5}, 031030 (2015).

\bibitem{lee_countermeasure_2016}
M.~S. Lee, B.~K. Park, M.~K. Woo, C.~H. Park, Y.-S. Kim, S.-W. Han
  \emph{et~al.}
\newblock Phys. Rev. A \textbf{94}, 062321 (2016).

\bibitem{vakhitov_tha_2011}
A.~Vakhitov, V.~Makarov, and D.~R. Hjelme.
\newblock J. Mod. Opt. \textbf{48}, 2023 (2001).

\bibitem{gisin_tha_2006}
N.~Gisin, S.~Fasel, B.~Kraus, H.~Zbinden, and G.~Ribordy.
\newblock Phys. Rev. A \textbf{73}, 022320 (2006).

\bibitem{stucki_pandp_2002}
D.~Stucki, N.~Gisin, O.~Guinnard, G.~Ribordy, and H.~Zbinden.
\newblock New J. Phys. \textbf{4}, 41 (2002).

\bibitem{yuan_resilience_2011_pra}
Z.~L. Yuan, J.~F. Dynes, and A.~J. Shields.
\newblock Appl. Phys. Lett. \textbf{98}, 231104 (2011).

\bibitem{AKS2017SPIE}
A.~M. Koehler-Sidki, J.~F. Dynes, M.~Lucamarini, G.~R. Roberts, A.~W. Sharpe,
  S.~J. Savory \emph{et~al.}
\newblock In \emph{Proc. SPIE 10442, Quantum Information Science and Technology
  III, 104420L} (2017).

\bibitem{lim_random_2015}
C.~C.~W. Lim, N.~Walenta, M.~Legr\'{e}, N.~Gisin, and H.~Zbinden.
\newblock IEEE Journal of Selected Topics in Quantum Electronics \textbf{21},
  192 (2015).

\bibitem{huang_testing_2016}
A.~Huang, S.~Sajeed, P.~Chaiwongkhot, M.~Soucarros, M.~Legr\'{e}, and
  V.~Makarov.
\newblock IEEE Journal of Quantum Electronics \textbf{52}, 1 (2016).

\bibitem{sajeed_watchdogdet_2015}
S.~Sajeed, I.~Radchenko, S.~Kaiser, J.-P. Bourgoin, A.~Pappa, L.~Monat
  \emph{et~al.}
\newblock Phys. Rev. A \textbf{91}, 032326 (2015).

\bibitem{lo12}
H.-K. Lo, M.~Curty, and B.~Qi.
\newblock Phys. Rev. Lett. \textbf{108}, 130503 (2012).

\bibitem{braunstein12}
S.~L. Braunstein and S.~Pirandola.
\newblock Phys. Rev. Lett. \textbf{108}, 130502 (2012).

\bibitem{dynes16}
J.~F. Dynes, S.~W.-B. Tam, A.~Plews, B.~Fr\"{o}hlich, A.~W. Sharpe,
  M.~Lucamarini \emph{et~al.}
\newblock Sci. Rep. \textbf{6}, 35149 (2016).

\bibitem{Yuan18}
Z.~Yuan, A.~Plews, R.~Takahashi, K.~Doi, W.~Tam, A.~Sharpe \emph{et~al.}
\newblock J. Lightwave Technol. \textbf{36}, 3427 (2018).

\bibitem{yuan_highspeedir_2007_pra}
Z.~L. Yuan, B.~E. Kardynal, A.~W. Sharpe, and A.~J. Shields.
\newblock Appl. Phys. Lett. \textbf{91}, 041114 (2007).

\bibitem{patel12a}
K.~A. Patel, J.~F. Dynes, A.~W. Sharpe, Z.~L. Yuan, R.~V. Penty, and A.~J.
  Shields.
\newblock Electron. Lett. \textbf{48}, 111 (2012).

\bibitem{comandar_gigahertz-gated_2015_pra}
L.~C. Comandar, B.~Fr\"{o}hlich, J.~F. Dynes, A.~W. Sharpe, M.~Lucamarini,
  Z.~L. Yuan \emph{et~al.}
\newblock J. Appl. Phys. \textbf{117}, 083109 (2015).

\bibitem{Comandar:rm_temp_det_pra}
L.~C. Comandar, B.~Fr\"{o}hlich, M.~Lucamarini, K.~A. Patel, A.~W. Sharpe,
  J.~F. Dynes \emph{et~al.}
\newblock Appl. Phys. Lett. \textbf{104}, 021101 (2014).

\bibitem{patel12prx}
K.~A. Patel, J.~F. Dynes, I.~Choi, A.~W. Sharpe, A.~R. Dixon, Z.~L. Yuan
  \emph{et~al.}
\newblock Phys. Rev. X \textbf{2}, 041010 (2012).

\bibitem{dynes08}
J.~F. Dynes, Z.~L. Yuan, A.~W. Sharpe, and A.~J. Shields.
\newblock Appl. Phys. Lett. \textbf{93}, 031109 (2008).

\bibitem{jiang_intrinsic_2013_pra}
M.-S. Jiang, S.-H. Sun, G.-Z. Tang, X.-C. Ma, C.-Y. Li, and L.-M. Liang.
\newblock Phys. Rev. A \textbf{88}, 062335 (2013).

\bibitem{yuan_avoiding_2010_pra}
Z.~L. Yuan, J.~F. Dynes, and A.~J. Shields.
\newblock Nat. Photonics \textbf{4}, 800 (2010).

\bibitem{thermal_Lydersen:10}
L.~Lydersen, C.~Wiechers, C.~Wittmann, D.~Elser, J.~Skaar, and V.~Makarov.
\newblock Opt. Express \textbf{18}, 27938 (2010).

\bibitem{fakedstate_2005}
V.~Makarov and D.~R. Hjelme.
\newblock J. Mod. Opt. \textbf{52}, 691 (2005).

\bibitem{KoehlerSidki18}
A.~Koehler-Sidki, J.~F. Dynes, M.~Lucamarini, G.~L. Roberts, A.~W. Sharpe,
  Z.~L. Yuan \emph{et~al.}
\newblock Phys. Rev. Applied \textbf{9}, 044027 (2018).

\bibitem{Koa06}
M.~Koashi.
\newblock Efficient quantum key distribution with practical sources and
  detectors (2006).
\newblock \eprint{arXiv:0609180}.

\bibitem{dixon_2017}
A.~R. Dixon, J.~F. Dynes, M.~Lucamarini, B.~Fr\"{o}hlich, A.~W. Sharpe,
  A.~Plews \emph{et~al.}
\newblock Sci. Rep. \textbf{7}, 1978 (2017).

\end{thebibliography}
\bibliographystyle{pra_bst_2}

\end{document}